\documentclass{aastex63}
\usepackage{lineno}
\usepackage{graphicx}
\usepackage{gensymb}

\usepackage{graphicx}
\usepackage{wrapfig}
\usepackage{sidecap}
\usepackage{float}
\accepted{ApJ Letters, July 3, 2020}
\shorttitle{NH$_3$-bearing species on Ariel}
\shortauthors{Cartwright et al.}

\begin{document}

\title{Evidence for ammonia-bearing species on the Uranian satellite Ariel \\ supports recent geologic activity}

\correspondingauthor{Richard J. Cartwright}
\email{rcartwright@seti.org}

\author[0000-0002-6886-6009]{Richard J. Cartwright$^a$}
\affiliation{The Carl Sagan Center at the SETI Institute \\
189 Bernardo Ave., Suite 200\\
Mountain View, CA 94043, USA}
\footnote{$^a$Visiting Astronomer at the Infrared Telescope Facility, which is operated by the University of Hawaii under contract NNH14CK55B with the National Aeronautics and Space Administration. }

\author{Chloe B. Beddingfield}
\affiliation{The Carl Sagan Center at the SETI Institute \\
	189 Bernardo Ave., Suite 200\\
	Mountain View, CA 94043, USA}
\affiliation{NASA-Ames Research Center
\\
	Mail Stop 245-1
\\
	Building N245, Room 204
\\
	P.O. Box 1
\\
	Moffett Field, CA 94035, USA}

\author{Tom A. Nordheim}
\affiliation{Jet Propulsion Laboratory \\
	4800 Oak Grove Dr.\\
	Pasadena, CA 91109, USA}

\author{Joe Roser}
\affiliation{The Carl Sagan Center at the SETI Institute \\
	189 Bernardo Ave., Suite 200\\
	Mountain View, CA 94043, USA}
\affiliation{NASA-Ames Research Center
\\
	Mail Stop 245-1
\\
	Building N245, Room 204
\\
	P.O. Box 1
\\
	Moffett Field, CA 94035, USA}

\author{William  M. Grundy}
\affiliation{Lowell Observatory \\
	1400 W Mars Hill Rd.\\
	Flagstaff, AZ 86001, USA}
\affiliation{Northern Arizona University \\
	S San Francisco St.\\
	Flagstaff, AZ 86011, USA}

\author{Kevin P. Hand}
\affiliation{Jet Propulsion Laboratory \\
	4800 Oak Grove Dr.\\
	Pasadena, CA 91109, USA}

\author{Joshua P. Emery}
\affiliation{Northern Arizona University \\
	S San Francisco St.\\
	Flagstaff, AZ 86011, USA}

\author{Dale P. Cruikshank}
\affiliation{NASA-Ames Research Center
\\
	Mail Stop 245-1
\\
	Building N245, Room 204
\\
	P.O. Box 1
\\
	Moffett Field, CA 94035, USA}

\author{Francesca Scipioni}
\affiliation{The Carl Sagan Center at the SETI Institute \\
	189 Bernardo Ave., Suite 200\\
	Mountain View, CA 94043, USA}



\begin{abstract}

We investigated whether ammonia-rich constituents are present on the surface of the Uranian moon Ariel by analyzing 32 near-infrared reflectance spectra collected over a wide range of sub-observer longitudes and latitudes. We measured the band areas and depths of a 2.2-$\micron$ feature in these spectra, which has been attributed to ammonia-bearing species on other icy bodies. Ten spectra display prominent 2.2-$\micron$ features with band areas and depths $>$ 2$\sigma$. We determined the longitudinal distribution of the 2.2-$\micron$ band, finding no statistically meaningful differences between Ariel's leading and trailing hemispheres, indicating that this band is distributed across Ariel's surface. We compared the band centers and shapes of the five Ariel spectra displaying the strongest 2.2-$\micron$ bands to laboratory spectra of various ammonia-bearing and ammonium-bearing species, finding that the spectral signatures of the Ariel spectra are best matched by ammonia-hydrates and flash frozen ammonia-water solutions. Our analysis also revealed that four Ariel spectra display 2.24-$\micron$ bands ($>$ 2$\sigma$ band areas and depths), with band centers and shapes that are best matched by ammonia ice. Because ammonia should be efficiently removed over short timescales by ultraviolet photons, cosmic rays, and charged particles trapped in Uranus' magnetosphere, the possible presence of this constituent supports geologic activity in the recent past, such as emplacement of ammonia-rich cryolavas and exposure of ammonia-rich deposits by tectonism, impact events, and mass wasting.

\end{abstract}

\keywords{Planetary surfaces --- 
Surface composition --- Surface processes --- Surface ices}


\section{Introduction} 

Planetesimals rich in ammonia (NH$_3$) were likely incorporated into Ariel and the other proto-classical moons as they formed in the Uranian subnebula \citep[e.g.,][]{lewis1972metal}. NH$_3$ is an efficient anti-freeze when mixed with liquid H$_2$O, which, if incorporated into Ariel's interior as a primordial constituent, would have helped this moon retain liquid H$_2$O in its subsurface for a much longer period of time compared to `pure' liquid H$_2$O, at temperatures as low as $\sim$176 K \citep[e.g.,][]{spohn2003oceans}. Analysis of crater densities suggests that some regions of Ariel's surface may be relatively young ($\sim$1 - 2 Ga, \citealt{zahnle2003cratering}). These regions display numerous examples of landforms indicative of resurfacing driven by tectonism and cryovolcanism \citep[e.g.,][]{smith1986voyager, schenk1991fluid, kargel1992ammonia, beddingfield2020Arielcryo}. Morphological assessment of potential cryovolcanic features in these regions points to emplacement of material sourced from Ariel's interior \citep{schenk1991fluid, beddingfield2020Arielcryo}. Furthermore, the estimated flow rheologies for potential cryovolcanic deposits on Ariel are consistent with emplacement of NH$_3$-rich cryolavas \citep[e.g.,][]{schenk1991fluid, kargel1992ammonia}. The geologic evidence therefore suggests that NH$_3$-bearing deposits sourced from Ariel's interior have played an important role in resurfacing this moon, in particular in regions estimated to be fairly young.

Supporting the geologic evidence for NH$_3$-rich deposits on Ariel, recent ground-based, near-infrared (NIR) observations have revealed a subtle absorption band centered near 2.2 $\micron$, consistent with the presence of NH$_3$-bearing species \citep{cartwright2018red}. Deposits rich in NH$_3$-hydrates and NH$_3$ ice, contained within the top meter of Ariel’s subsurface, are likely decomposed by magnetospheric charged particle bombardment over geologically short timescales \citep{moore2007ammonia}, suggesting replenishment of NH$_3$ from Ariel's interior in the recent past. However, the spatial distribution of the 2.2-$\micron$ band was not previously assessed, limiting our ability to investigate the origin of this feature. Furthermore, the species contributing to the 2.2-$\micron$ band have not been assessed, and both NH$_3$-hydrates and NH$_4$-rich salts could contribute to this feature. The presence of these constituents could have important implications for Ariel’s geologic history and the evolution of its surface composition. In this work, we investigated the longitudinal distribution and composition of the 2.2-$\micron$ band, conducting band parameter measurements on 32 NIR spectra, which confirm the presence of the 2.2-$\micron$ band on Ariel. We also investigated the possible presence of a 2.24-$\micron$ band, which could result from NH$_3$ ice.

\section{Data and Methods}

\subsection{Observations and Data Reduction} 
The nine new reflectance spectra reported here were collected between 2017 and 2019 using the NIR SpeX spectrograph/imager at NASA's Infrared Telescope Facility (IRTF), operating in low resolution PRISM mode and moderate resolution short cross-dispersed (SXD) mode \citep[e.g.,][]{rayner2003spex}. These IRTF/SpeX observations were made using `AB' nodding, where the target is observed in two different positions along the 15'' slit, separated by 7.5''. The resulting `A' and `B' exposures are separated into sequential pairs, and the `B' exposures are subtracted from the `A' exposures to perform first order sky emission correction. Calibration frames (flat fields and arc lamps) were generated using quartz and argon lamps with an internal integrating sphere. Data extraction and calibration were conducted using custom software and the Spextool data reduction suite \citep{cushing2004spextool}. Extracted spectra from each night were co-added during reduction to boost signal-to-noise (S/N).  Ariel spectra were divided by solar analog star spectra, observed multiple times on the same nights, to remove the solar spectrum and provide additional atmospheric correction. All star spectra were collected within $\pm 0.1$ airmass of the Ariel spectra. The analog stars we observed between 2017 and 2019 were: HD 3628, HD 11532, HD 12124, and BD+09 213. The 23 previously reported Ariel spectra were also collected using IRTF/SpeX between 2000 and 2016 \citep{grundy2003discovery, grundy2006distributions, cartwright2015distribution, cartwright2018red}. We refer the reader to these papers for detailed descriptions of the data reduction routines utilized by each team. Observation details for all 32 Ariel spectra are summarized in Table 1. Each team used different slit widths (0.3'', 0.5'', and 0.8'') with SpeX to achieve a variety of observing goals. The 2.2-$\micron$ band can be detected in both PRISM and SXD mode using all slit widths reported here. 

\subsection{Band Parameter Analyses} 
The central wavelength of the 2.2-$\micron$ band can vary between $\sim$2.20 and 2.22 $\micron$ in spectra of Ariel (Figure 1). Furthermore, the wavelength range covered by the 2.2-$\micron$ band varies between 2.18 and 2.23 $\micron$, with band widths of $\sim$0.03 to 0.05 $\micron$. To assess the spatial distribution and spectral signature of the 2.2-$\micron$ band, we measured the band area and band depth of this feature in each of the 32 Ariel spectra, utilizing a custom data processing program that our team has used previously to conduct band parameter analyses of other icy constituents detected on the Uranian moons  \citep{cartwright2015distribution,cartwright2018red}. The program ingested individual spectra and fit their continua between 2.17 to 2.24 $\micron$, spanning the entire wavelength range covered by the 2.2-$\micron$ band. To simulate the continuum of each spectrum (i.e., continua without 2.2-$\micron$ or 2.24-$\micron$ absorption bands), we generated synthetic spectra using Hapke-based radiative transfer models (e.g., \citet{hapke2012theory}, Appendix 6.1). These synthetic spectra were generated using laboratory measured optical constants for the primary constituents that have been detected previously on Ariel: H$_2$O ice \citep{mastrapa2008optical}, amorphous carbon \citep{rouleau1991shape}, and CO$_2$ ice \citep{hansen1997spectral}. The program then divided each Ariel spectrum by its modeled continuum between 2.17 to 2.24 $\micron$, and measured the areas of the resulting continuum-divided bands using the trapezoidal rule. To estimate uncertainties, the program utilized Monte Carlo simulations that resample the 1$\sigma$ errors of each spectral channel within the wavelength range of each band (iterated 20,000 times). 

To measure the 2.2-$\micron$ feature depths, we first assigned the wavelength at the deepest part of each continuum-divided feature as the band center. The program then averaged the reflectance for all spectral channels within $\pm 0.004$ $\micron$ of these band centers, thereby calculating a mean reflectance for each band center. To calculate the 2.2-$\micron$ band depths, the program subtracted these mean reflectances from 1. To estimate the uncertainties of each band depth measurement, the program added the 1$\sigma$ errors of all spectral channels included in the mean reflectance measurement, in quadrature, and divided by the number of channels (n), thereby calculating the mean uncertainty ($\overline{\sigma}$). The program then calculated the standard deviation of the mean ($\sigma$$\overline{_x}$ = $\sigma$/$\sqrt{n}$) to estimate the point-to-point variation for each band depth measurement and calculated the final band depth error by summing $\overline{\sigma}$ and $\sigma$$\overline{_x}$ in quadrature. We show an illustrative example of our band area and depth measurement procedure in Appendix 6.2.

We also conducted band parameter measurements of the 2.24-$\micron$ band, using the same procedures described above for the 2.2-$\micron$ band. The 2.24-$\micron$ band spans $\sim$2.22 to 2.26 $\micron$ (Figure 1), slightly overlapping the wavelength range of the 2.2-$\micron$ band, with band centers between $\sim$2.24 and 2.245 $\micron$ (continua spanning 2.21 to 2.27 $\micron$).

\begin{table}[]
	\caption {IRTF/SpeX observations of Ariel.} 
	\hskip-0.8cm\begin{tabular}{*9c}
		\hline\hline
		\begin{tabular}[c]{@{}l@{}}\hspace{-1 cm}Sub-\\  \hspace{-1 cm}Observer\\  \hspace{-1 cm}Long. ($\degree$)\end{tabular} & \begin{tabular}[c]{@{}l@{}} \hspace{-1 cm}Sub-\\  \hspace{-1 cm}Observer \\  \hspace{-1 cm}Lat.  ($\degree$)\end{tabular} & UT Date & \begin{tabular}[c]{@{}l@{}} \hspace{-1 cm}UT Time  \\  \hspace{-1 cm}(mid-expos)\end{tabular} & \begin{tabular}[c]{@{}l@{}} \hspace{-1 cm}Integration \\ \hspace{-1 cm}Time   (min)\end{tabular} & \begin{tabular}[c]{@{}l@{}} \hspace{-1 cm}SpeX \\  \hspace{-1 cm}Observing \\  \hspace{-1 cm}Mode\end{tabular} & \begin{tabular}[c]{@{}l@{}} \hspace{-1 cm}Slit \\  \hspace{-1 cm}Width \\  \hspace{-1 cm}('')\end{tabular} & \begin{tabular}[c]{@{}l@{}} \hspace{-1 cm}Average \\  \hspace{-1 cm}Resolving \\  \hspace{-1 cm}Power ($\lambda$/$\Delta$$\lambda$)\end{tabular} & References \\
		\hline
		9.2 & 31.8 & 9/18/15 & 10:05 & 7.5 & PRISM & 0.8 & 93.8 & Cartwright et al. (2018) \\
		15.3 & 27.8 & 9/15/14 & 11:35 & 88 & SXD & 0.8& 750 & Cartwright et al. (2018) \\
		38.8 & 35.8 & 9/20/16 & 14:15 & 7.5 & PRISM & 0.8& 93.8 & Cartwright et al. (2018) \\
		53.6 & -16.0 & 8/9/03 & 12:15 & 156 & SXD & 0.3& 2000 & Grundy et al. (2006) \\
		79.8 & -19.4 & 7/17/02 & 13:25 & 108 & SXD & 0.5& 1200 & Grundy et al. (2003) \\
		87.8 & 24.0 & 9/5/13 & 11:10 & 92 & SXD & 0.8& 750 & Cartwright et al. (2015) \\
		93.5 & -18.1 & 10/4/03 & 5:45 & 108 & SXD & 0.3& 2000 & Grundy et al. (2006) \\
		110.1 & 32.0 & 9/11/15 & 13:30 & 44 & SXD & 0.8& 750 & Cartwright et al. (2018) \\
		132.2 & 28.5 & 8/24/14 & 14:05 & 40 & SXD & 0.8& 750 & Cartwright et al. (2018) \\
		137.6 & 34.6 & 10/21/16 & 12:40 & 7.5 & PRISM & 0.8& 93.8 & Cartwright et al. (2018) \\
		144.8 & 43.2 & 10/12/18 & 9:30 & 73 & SXD & 0.5& 1200 & This work \\
		159.9 & -11.1 & 7/15/04 & 12:00 & 112 & SXD & 0.3& 2000 & Grundy et al. (2006) \\
		200.0 & -15.9 & 8/5/03 & 12:00 & 84 & SXD & 0.3& 2000 & Grundy et al. (2006) \\
		201.3 & 33.2 & 1/23/17 & 5:45 & 10.5 & PRISM & 0.8& 93.8 & Cartwright et al. (2018) \\
		205.5 & 46.1 & 11/7/19 & 11:20 & 60 & SXD & 0.5& 1200 & This work \\
		219.8 & -17.2 & 9/7/03 & 9:35 & 90 & SXD & 0.3& 2000 & Grundy et al. (2006) \\
		224.8 & 31.8 & 9/17/15 & 9:40 & 10 & PRISM & 0.8& 93.8 & Cartwright et al. (2018) \\
		233.6 & 32.0 & 9/12/15 & 10:15 & 8 & PRISM & 0.8& 93.8 & Cartwright et al. (2018) \\
		233.8 & -23.1 & 7/5/01 & 14:10 & 50 & SXD & 0.5& 1200 & Grundy et al. (2003) \\
		242.0 & 42.2 & 11/7/18 & 6:50 & 10.5 & PRISM & 0.8& 93.8 & This work \\
		244.8 & 47.1 & 10/18/19 & 14:05 & 7.5 & PRISM & 0.8& 93.8 & This work \\
		253.9 & 29.2 & 12/2/15 & 5:25 & 9 & PRISM & 0.8& 93.8 & Cartwright et al. (2018) \\
		257.6 & -29.5 & 9/6/00 & 7:35 & 76 & SXD & 0.8& 750 & Cartwright et al. (2015) \\
		268.3 & 39.0 & 10/15/17 & 8:00 & 40 & SXD & 0.5& 1200 & This work \\
		273.2 & 42.2 & 11/7/18 & 12:00 & 42 & SXD & 0.5& 1200 & This work \\
		278.3 & 24.8 & 8/7/13 & 13:20 & 44 & SXD & 0.8& 750 & Cartwright et al. (2015) \\
		294.8 & 39.5 & 9/30/17 & 9:30 & 120 & SXD & 0.5& 1200 & This work \\
		294.8 & -19.3 & 7/16/02 & 13:10 & 140 & SXD & 0.5& 1200 & Grundy et al. (2003) \\
		297.0 & 39.2 & 10/10/17 & 11:50 & 12.5 & PRISM & 0.8& 93.8 & This work \\
		304.8 & -23.2 & 7/8/01 & 14:40 & 48 & SXD & 0.5& 1200 & Grundy et al. (2003) \\
		316.6 & -18.2 & 10/8/03 & 7:55 & 132 & SXD & 0.3& 2000 & Grundy et al. (2006) \\
		334.4 & 39.7 & 9/25/17 & 15:10 & 7.5 & PRISM & 0.8& 93.8 & This work \\
		\hline
	\end{tabular} 
\end{table}

\section{Results and Analysis} 

\begin{table}[]
	\caption {Measurements of the 2.2-$\micron$ band.} 
	\begin{tabular}{*7c}
		\hline\hline
		\begin{tabular}[c]{@{}l@{}}\hspace{-1 cm}Sub-\\  \hspace{-1 cm}Observer \\ \hspace{-1 cm}Long. ($\degree$)\end{tabular} & \begin{tabular}[c]{@{}l@{}} \hspace{-1 cm}Sub-\\  \hspace{-1 cm}Observer \\  \hspace{-1 cm}Lat.  ($\degree$)\end{tabular}  & \begin{tabular}[c]{@{}l@{}} \hspace{-1 cm}Wavelength \\  \hspace{-1 cm}Range ($\micron$) \end{tabular} &
		\begin{tabular}[c]{@{}l@{}} \hspace{-1 cm}Band Area \\ \hspace{-1 cm}(10$^-$$^4$ $\micron$)   \end{tabular} & \begin{tabular}[c]{@{}l@{}} \hspace{-1 cm}Band Depth \\  \hspace{-1 cm}($\micron$)\end{tabular} & \begin{tabular}[c]{@{}l@{}} \hspace{-1 cm}$>$ 2$\sigma$ Band \\  \hspace{-1 cm}Area $\&$ Depth \\  \hspace{-1 cm}Measurements? \end{tabular} & \begin{tabular}[c]{@{}l@{}} \hspace{-1 cm}Band \\  \hspace{-1 cm}Center\\   \hspace{-1 cm}($\micron$) \end{tabular}\\
		\hline
		9.2 & 31.8 & 2.190 - 2.230 & 1.11 $\pm$  2.43 & 0.012 $\pm$  0.016 & No & -  \\
		15.3 & 27.8 & 2.198 - 2.228 & 2.01 $\pm$  0.43 & 0.011 $\pm$  0.004 & Yes  & 2.214 \\
		38.8 & 35.8 & 2.180 - 2.220 & 1.68 $\pm$  1.28 & 0.002 $\pm$  0.007 & No  & -  \\
		53.6 & -16.0 & 2.180 - 2.220 & 6.84 $\pm$  1.28 & 0.032 $\pm$  0.008 & Yes  & 2.202 \\
		79.8 & -19.4& 2.191 - 2.225 & 3.79 $\pm$  0.67 & 0.023 $\pm$  0.005 & Yes  & 2.209 \\
		87.8 & 24.0 &  2.190 - 2.225 & 2.64 $\pm$  0.54 & 0.010 $\pm$  0.005 & No  & -  \\
		93.5 & -18.1 & 2.190 - 2.225 & 1.91 $\pm$  1.10 & 0.012 $\pm$  0.007 & No & - \\
		110.1 & 32.0 & 2.190 - 2.230 & 0.94 $\pm$  0.59 & 0.005 $\pm$  0.005 & No  & - \\
		132.2 & 28.5 & 2.180 - 2.220 & 0.43 $\pm$  0.65 & 0.003 $\pm$  0.004 & No   & - \\
		137.6 & 34.6 &  2.180 - 2.220 & 1.21 $\pm$  1.64 & 0.011 $\pm$ 0.019 & No  & -  \\
		144.8 & 43.2 & 2.190 - 2.232 & 8.30 $\pm$ 0.80 & 0.032 $\pm$  0.006 & Yes & 2.209 \\
		159.9 & -11.1 & 2.191 - 2.230 & 6.75 $\pm$  1.12 & 0.024 $\pm$  0.007 & Yes & 2.209 \\
		200.0 & -15.9 & 2.180 - 2.220 & 1.19 $\pm$  1.14 & 0.014 $\pm$  0.009 & No & - \\
		201.2 & 33.2 & 2.180 - 2.220 & 0.36 $\pm$ 1.54 & 0.001 $\pm$  0.016 & No  & - \\
		205.5 & 46.1 & 2.180 - 2.220 & 0.70 $\pm$  0.42 & 0.003 $\pm$  0.004 & No & -  \\
		219.8 & -17.2 & 2.180 - 2.220 & -1.45 $\pm$  0.82 & 0.000 $\pm$  0.005 & No  & -  \\
		224.8 & 31.8 & 2.200 - 2.230 & 0.96 $\pm$ 0.73 & 0.000 $\pm$  0.006 & No & -  \\
		233.6 & 32.0 & 2.177 - 2.222 & 3.72 $\pm$  1.27 & 0.019 $\pm$  0.008 & Yes & 2.202 \\
		233.8 & -23.1 & 2.180 - 2.220 & 1.28 $\pm$  1.84 & 0.020 $\pm$  0.010 & No  & - \\
		242.0 & 42.2 & 2.187 - 2.227 & 2.35 $\pm$  0.93 & 0.008 $\pm$  0.007 & No  & -  \\
		244.8 & 47.1 & 2.187 - 2.227 & -1.50 $\pm$  1.35 & 0.001 $\pm$  0.013 & No  & -  \\
		253.9 & 29.2 & 2.200 - 2.230 & -0.47 $\pm$  0.75 & 0.002 $\pm$ 0.070 & No & - \\
		257.6 & -29.5 & 2.183 - 2.205 & 2.44 $\pm$  0.60 & 0.019 $\pm$  0.007 & Yes  & 2.198 \\
		268.3 & 39.0 & 2.192 - 2.222 &  1.55 $\pm$  0.66 & 0.016 $\pm$  0.006 & Yes  & 2.214 \\
		273.2 & 42.2 & 2.192 - 2.222 & 0.58 $\pm$  1.50 & 0.014 $\pm$  0.012 & No  & - \\
		278.3 & 24.8 & 2.192 - 2.222 & -1.08 $\pm$  1.43 & -0.007 $\pm$  0.012 & No  & - \\
		294.8 & 39.5 & 2.192 - 2.222 & 2.54 $\pm$  0.73 & 0.007 $\pm$  0.006 & No  & - \\
		294.8 & -19.3 & 2.192 - 2.222 & 2.46 $\pm$  0.52 & 0.012 $\pm$  0.006 & No  & -  \\
		297.0 & 39.2 & 2.187 - 2.227 & 1.70 $\pm$  1.55 & 0.001 $\pm$  0.014 & No  & - \\
		304.8 & -23.2 & 2.196 - 2.226 & 2.58 $\pm$  1.87 & 0.013 $\pm$  0.009 & No  & - \\
		316.6 & -18.2 & 2.192 - 2.222 & 5.48 $\pm$  0.98 & 0.038 $\pm$  0.010 & Yes & 2.203 \\
		334.4 & 39.7 & 2.185 - 2.220 & 1.63 $\pm$  0.74 & 0.023 $\pm$  0.008 & Yes  & 2.200  \\
		\hline
	\end{tabular}
\end{table}

\subsection{IRTF/SpeX Spectra of Ariel} 
We report nine new disk-integrated spectra of Ariel, collected with IRTF/SpeX operating in PRISM and SXD mode (Appendix 6.3). All nine of these spectra display the 1.52-$\micron$ and 2.02-$\micron$ H$_2$O ice bands detected in previously collected NIR spectra of this moon \citep[e.g.,][]{cruikshank1981uranian,grundy2003discovery,grundy2006distributions,cartwright2015distribution,cartwright2018red}. Furthermore, the five new spectra collected in SXD mode display clear evidence for the narrow CO$_2$ ice bands detected previously between 1.9 and 2.1 $\micron$  \citep{grundy2003discovery,grundy2006distributions, cartwright2015distribution,cartwright2018red}. Analysis of these H$_2$O and CO$_2$ ice absorption features is beyond the scope of this paper and will be included in future work.

\subsection{Band parameter analyses} 
We assessed the presence of the 2.2-$\micron$ band by conducting band area and depth measurements, which we report in Table 2 (along with their 1$\sigma$ uncertainties). Our analysis determined that 10 of the 32 Ariel spectra have 2.2-$\micron$ bands with area and depth measurements that are both $>$ 2$\sigma$. The band centers for these 10 spectra cluster around three wavelength intervals: 2.198 to 2.203 $\micron$, 2.209 $\micron$, and 2.214 $\micron$ (Figure 1). Evidence for the 2.2-$\micron$ band is weaker in the other 22 spectra, making robust assignment of their band centers challenging, and consequently, we do not report band centers for these data. Along with the 2.2-$\micron$ band, four spectra of Ariel also display 2.24 $\micron$ bands with $>$ 2$\sigma$ band areas and depths (Table 3 in Appendix 6.4). In total, 12 different Ariel spectra display 2.2-$\micron$ and/or 2.24-$\micron$ bands (Figure 1).

\begin{figure}[h]
	\includegraphics[scale=0.69]{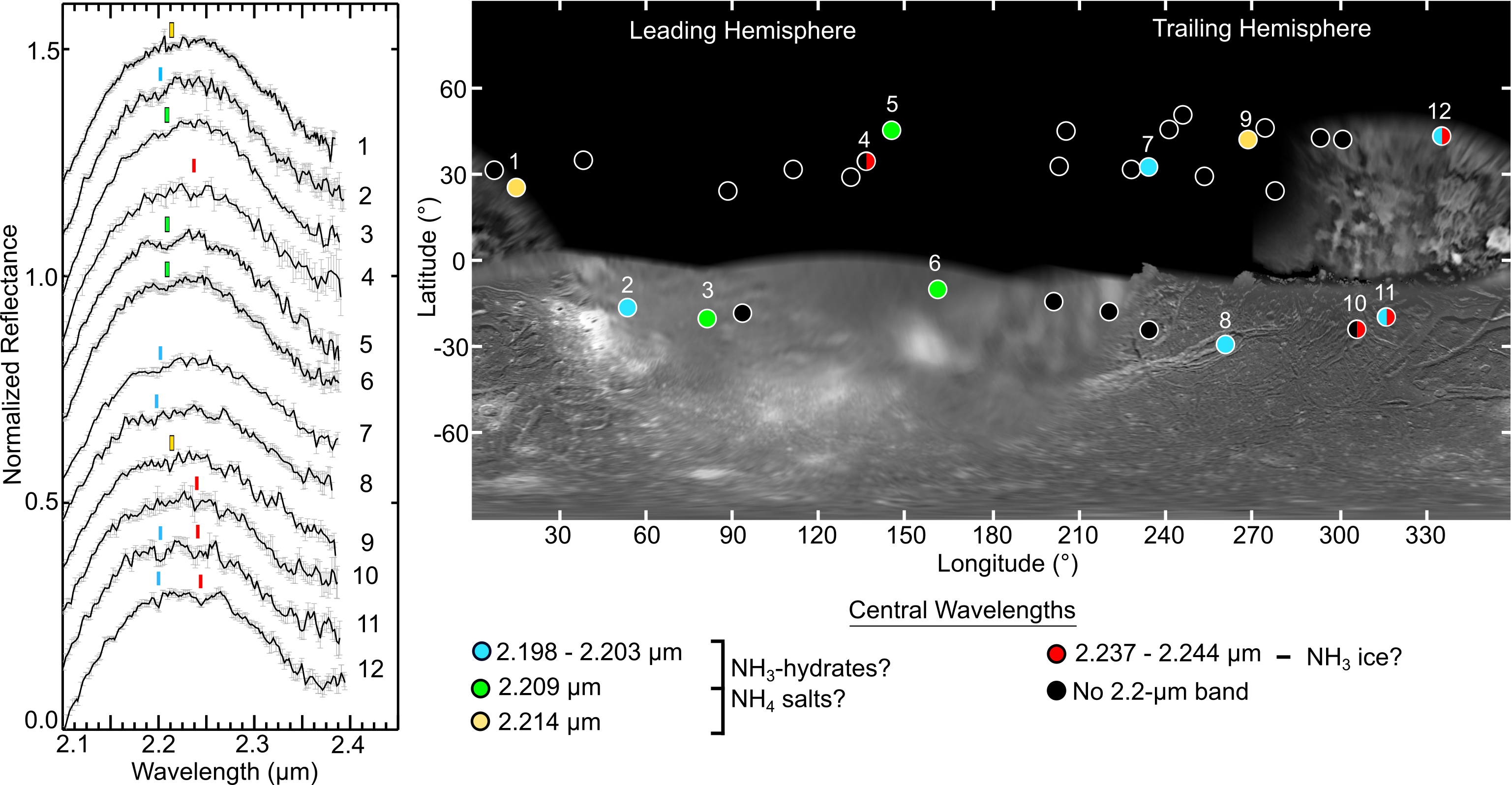}
	\caption{\textit{Left: The 12 IRTF/SpeX spectra of Ariel displaying 2.2-$\micron$ and/or 2.24-$\micron$ bands with $>$ 2$\sigma$ band areas and depths (Table 2), offset vertically for clarity and numbered 1 through 12. The 1$\sigma$ uncertainties for each spectrum are shown in light gray. These spectra have been lightly smoothed using a binning routine with a 3 to 10 pixel-wide window. The central wavelengths of the detected 2.2-$\micron$ bands are located: between 2.198 and 2.203 $\micron$ (blue markers), at 2.209 $\micron$ (green markers), and at 2.214 $\micron$ (yellow markers). The central wavelength of the 2.24-$\micron$ band is between 2.237 and 2.244 $\micron$ (red markers). Right: Voyager 2/Imaging Science System image mosaic of Ariel (courtesy NASA/JPL/Caltech/USGS, http://maps.jpl.nasa.gov/uranus.html), with night-side sections illuminated by `Uranus shine' \citep{stryk2008voyager}. The mid-observation longitudes and latitudes for all 32 Ariel spectra are indicated with dots that represent the center of the target disk at the time of each observation (each collected spectrum averages over an entire hemisphere). The 12 spectra shown on the left are indicated by color-filled dots (numbered 1 through 12). Spectra that do not display 2.2-$\micron$ bands are shown as black-filled dots (i.e., spectra with band areas and depths $<$ 2$\sigma$). For continuum-divided examples of the 2.2-$\micron$ and 2.24-$\micron$ bands, please see Figure 3. NH$_3$- and NH$_4$-rich constituents that could be contributing to these bands are investigated in Section 3.4.}}\vspace{0.1 cm}
\end{figure}

\subsection{Spatial distribution of the 2.2-$\micron$ band} 
Previous work has demonstrated that Ariel displays clear longitudinal trends in its surface composition and photometric properties, with significantly stronger ($>$ 3$\sigma$) H$_2$O ice bands on its leading hemisphere (1 - 180$\degree$ longitude) and stronger CO$_2$ ice bands and higher NIR geometric albedos ($>$ 3$\sigma$) on its trailing hemisphere (181 - 360$\degree$ longitude) \citep[e.g.,][]{grundy2006distributions,cartwright2020probing}. Consequently, we searched for similar longitudinal trends in the distribution of the 2.2-$\micron$ band on Ariel by calculating the mean 2.2-$\micron$ band areas for Ariel's leading and trailing hemispheres and comparing them: 3.13 $\pm$ 0.84 $\micron$ and 1.44 $\pm$ 0.48 $\micron$, respectively. We also calculated and compared the mean 2.2-$\micron$ band depths for Ariel's leading and trailing hemispheres: 0.015 $\pm$ 0.004 $\micron$ and 0.010 $\pm$ 0.003 $\micron$, respectively. Comparison of these values demonstrates that Ariel's mean 2.2-$\micron$ band areas are slightly stronger on its leading hemisphere but not at a statistically meaningful level ($<$ 2$\sigma$). Similarly, Ariel's mean 2.2-$\micron$ band depths display essentially no difference ($<$ 1$\sigma$) between its leading and trailing hemispheres.

As another test, we plotted the 32 individual 2.2-$\micron$ band area and depth measurements as a function of longitude, and fit them with a mean and sinusoidal model to determine whether the 2.2-$\micron$ band displays leading/trailing trends in its distribution (Figure 2). The mean model represents a surface displaying no longitudinal trends in the distribution of the 2.2-$\micron$ band. Conversely, the sinusoidal model represents a surface with significant differences in the longitudinal distribution of the 2.2-$\micron$ band (i.e., stronger 2.2-$\micron$ bands on Ariel's leading or trailing hemisphere). We compared these model fits using an \textit{F}-test \citep[e.g.,][]{lomax2013introduction}, with a null hypothesis that there is no meaningful difference between these two model fits. The results of the \textit{F}-test demonstrate that there is no statistically significant difference between the mean and sinusoidal models for either the band area or depth measurements (Table 4 in Appendix 6.5). 

Thus, comparison of the mean 2.2-$\micron$ band areas and depths, and \textit{F}-test analysis of the individual band parameter measurements, indicates that there are no meaningful trends in the longitudinal distribution of the 2.2-$\micron$ band across the surface of Ariel, unlike the longitudinal distribution of H$_2$O and CO$_2$ ice on this moon. The small number of Ariel spectra displaying a 2.24-$\micron$ band (four) prevents similar quantitative analysis of the longitudinal distribution of this feature. Nevertheless, we note that three of the four detected 2.24-$\micron$ bands are clustered between longitudes 305$\degree$ to 335$\degree$ on Ariel's trailing hemisphere, which is dominated by large chasmata, as well as other landforms suggestive of geologic activity in the recent past. The fourth 2.24-$\micron$ band (137.6$\degree$ longitude) is proximal to two of the strongest 2.2-$\micron$ bands detected on Ariel (Spectra 5 and 6 in Figure 1).   

\begin{figure}[h]
	\includegraphics[scale=0.65]{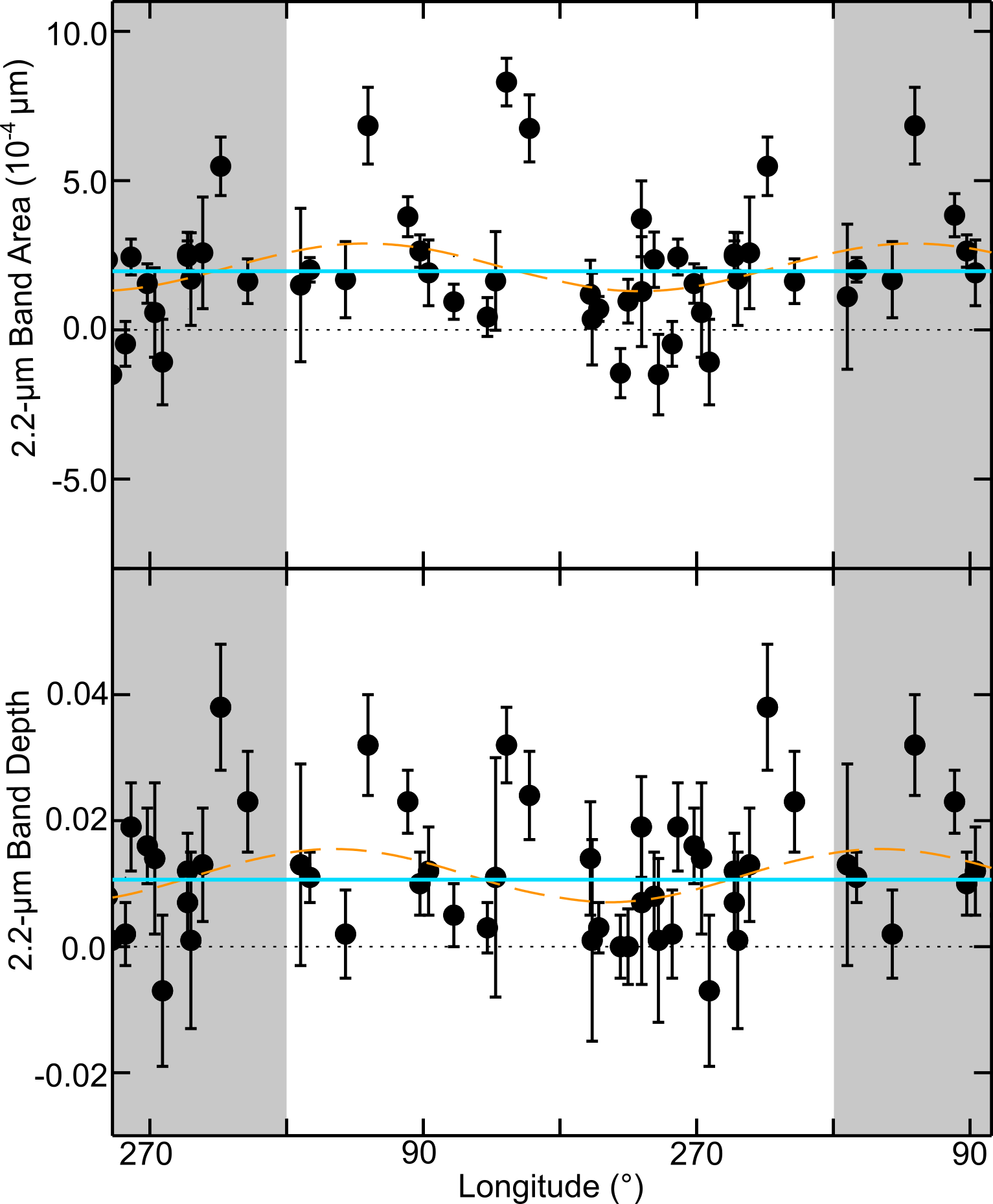}
	\caption{\textit{2.2-$\micron$ band area (top plot) and band depth (bottom plot) measurements and 1$\sigma$ uncertainties for all 32 Ariel spectra, shown as a function of sub-observer longitude (Table 2). The dashed orange lines on both plots are sinusoidal fits to the data, and the solid blue lines are the mean measurements. The maxima of the sinusoidal fits are free parameters and are not locked to specific longitudes. Duplicate longitudes are shown as gray-toned regions. The sinusoidal and mean model fits were compared using an \textit{F}-test, which determined that there is no statistically meaningful difference (p $<$ 0.05) between these two model fits for either the band area or depth measurements, with \textit{p} values of $\sim$0.09 and $\sim$0.10, respectively (\textit{F}-test results summarized in Table 4 in Appendix 6.5). These results indicate that there are no leading/trailing hemispherical asymmetries in the distribution of the 2.2-$\micron$ band. Several of the individual data points are significantly larger than the mean band area and band depth values (between longitudes $\sim$80 to 160$\degree$ and $\sim$315 to 345$\degree$), suggesting that these two regions of Ariel's surface might have larger concentrations of the constituents contributing to the 2.2-$\micron$ band. }}\vspace{0.1 cm}
\end{figure} 

\begin{figure}[h]
	\includegraphics[scale=0.75]{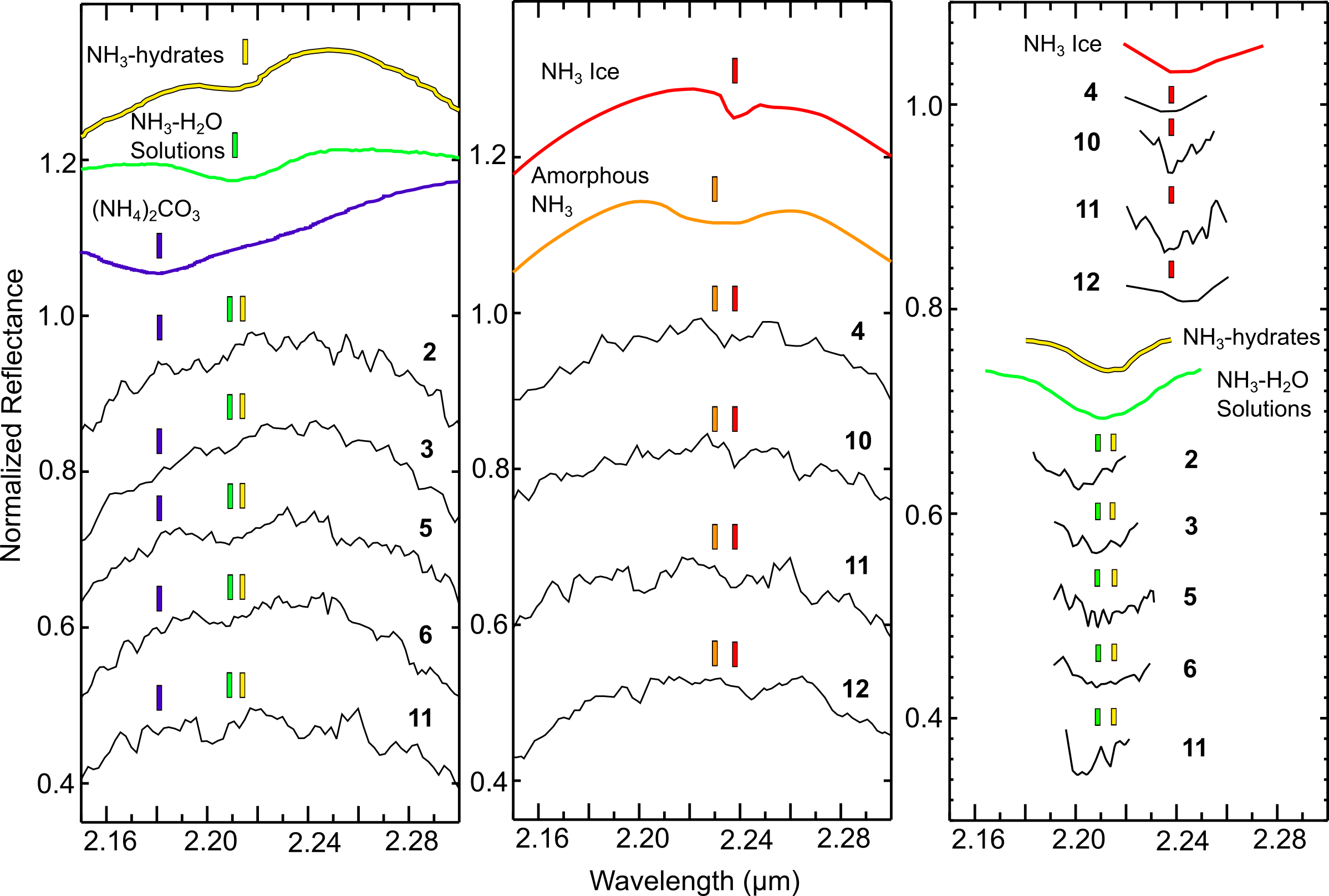}
	\caption{\textit{Left: Ariel spectra with $>$ 3$\sigma$ 2.2-$\micron$ band measurements (black lines) compared to laboratory spectra of mixed NH$_3$-hydrates (yellow; \citet{brown1988search}), NH$_3$-H$_2$O solutions (green; Tom Nordheim, private communication), and ammonium carbonate ((NH$_4$)$_2$CO$_3$, purple; \citet{berg2016reflectance}), offset vertically for clarity. Middle:  Ariel spectra with $>$ 2$\sigma$ 2.24-$\micron$ band measurements (black lines) compared to synthetic spectra that include NH$_3$ ice (red; \citet{sill1980absorption}) and amorphous NH$_3$ (orange; \citet{roser2018laboratory}), offset vertically for clarity. Right: Continuum-divided 2.2-$\micron$ and 2.24-$\micron$ bands detected on Ariel compared to continuum-divided 2.2-$\micron$ and 2.24-$\micron$ bands from laboratory spectra of NH$_3$ ice (red), NH$_3$-hydrates (yellow), and NH$_3$-H$_2$O solutions (green), offset vertically for clarity. In each plot, the colored markers indicate the central wavelength of the 2.2-$\micron$ and 2.24-$\micron$ bands measured in the laboratory spectra. All Ariel spectra have been lightly smoothed using a binning routine with a 3 to 10 pixel wide window. The Ariel spectra are numbered using the same sequence shown in Figure 1. }} \vspace{0.1 cm}
\end{figure}

\subsection{Comparing Ariel spectra to laboratory spectra of NH$_3$- and NH$_4$-bearing species} 
To investigate the composition of the constituents contributing to the detected 2.2-$\micron$ and 2.24-$\micron$ bands, we compared the Ariel data to reflectance spectra of different NH$_3$-bearing species measured in the laboratory, including: mixtures of multiple NH$_3$-hydrates \citep[e.g.,][]{brown1988search},  flash frozen NH$_3$-H$_2$O solutions (Tom Nordheim, private communication), and ammonium carbonate ((NH$_4$)$_2$CO$_3$) \citep{berg2016reflectance}. Although other NH$_4$ salts display similar 2.2-$\micron$ bands to (NH$_4$)$_2$CO$_3$, we selected this particular constituent for comparison because of the large abundance of CO$_2$ ice on Ariel, which could chemically interact with NH$_3$ and other constituents due to magnetospheric charged particle bombardment of Ariel's surface. We refer the reader to \cite{berg2016reflectance} for more information on a wide variety of NH$_4$-bearing constituents. We also compared the Ariel data to Hapke-based synthetic spectra (Appendix 6.1), generated using optical constants for NH$_3$ ice \citep{sill1980absorption} and amorphous NH$_3$ \citep{roser2018laboratory}, mixed with H$_2$O ice, CO$_2$ ice, and amorphous C. 

Because the detected 2.2-$\micron$ features are fairly weak and display morphological differences between the different Ariel spectra (i.e., variable band centers, areas, and depths), we focused our comparison on the five Ariel spectra displaying the strongest 2.2-$\micron$ bands ($>$ 3$\sigma$ band areas and depths) (Figure 3). This comparison demonstrates that the 2.2-$\micron$ bands in the spectra collected at mid-observation longitudes 79.8$\degree$, 144.8$\degree$, and 159.9$\degree$ (Spectra 3, 5, and 6, respectively,  in Figures 1 and 3) have band shapes and central wavelengths that are remarkably consistent with the laboratory spectra of NH$_3$-H$_2$O solutions (2.210 $\micron$) and NH$_3$-hydrates (2.215 $\micron$). The 2.2-$\micron$ bands in the spectra collected at mid-observation longitudes 53.6$\degree$ and 316.6$\degree$ (Spectra 2 and 11, respectively, in Figures 1 and 3) have band centers shifted to shorter wavelengths, possibly consistent with (NH$_4$)$_2$CO$_3$ (2.181 $\micron$), or perhaps other NH$_4$-bearing salts. Albeit, the shape of the 2.2-$\micron$ band in these two Ariel spectra is not a close match to the shape of the spectral continuum of (NH$_4$)$_2$CO$_3$. Three of the four spectra displaying 2.24-$\micron$ bands have band centers that are close matches to NH$_3$ ice (2.238 $\micron$), while the fourth spectrum (Spectrum 12 in Figures 1 and 3) has a 2.24-$\micron$ band shifted to slightly longer wavelengths. Amorphous NH$_3$ does not provide a good match to either the 2.2-$\micron$ or 2.24-$\micron$ band. This mismatch is perhaps unsurprising given that amorphous NH$_3$ transitions to polycrystalline ice at ~65 K \citep{dawes2007morphological}, and is therefore unstable at Ariel's estimated peak surface temperatures ($\sim$80 to 90 K near the subsolar point, \citealt{sori2017wunda}). Thus, laboratory reflectance spectra of NH$_3$-H$_2$O solutions and NH$_3$-hydrates, and synthetic spectra including NH$_3$ ice, represent the best matches to the 2.2-$\micron$ and 2.24-$\micron$ bands detected in the Ariel spectra.

\section{Discussion and Conclusions} 
Our analysis demonstrates that a 2.2-$\micron$ band is present at the $>$ 2$\sigma$ level in ten spectra collected over a wide range of sub-observer longitudes and latitudes on Ariel. We find no leading/trailing hemispherical trends in the distribution of the 2.2-$\micron$ band, unlike the distribution of CO$_2$ and H$_2$O ice on this moon \citep[e.g.,][]{grundy2006distributions}. The longitudinal distributions of CO$_2$ and H$_2$O are likely controlled by exogenic processes that modify the surface chemistry of Ariel, in particular charged particle radiolysis and dust impacts \citep{grundy2003discovery, grundy2006distributions, cartwright2015distribution, cartwright2018red}. The lack of spatial trends in the distribution of the 2.2-$\micron$ band supports an origin from geologic landforms distributed across Ariel's surface. Impact craters, tectonic faults and fractures, potential cryovolcanic constructs, and mass wasting features have been identified across Ariel, and these landforms could represent suitable sources for NH$_3$-bearing constituents originating in the interior of this moon.

The 2.2-$\micron$ and 2.24-$\micron$ bands have spectral signatures that are similar to NH$_3$-H$_2$O solutions, NH$_3$-hydrates, and NH$_3$ ice. The Ariel spectra we analyzed, however, do not display consistent 2.2-$\micron$ and 2.24-$\micron$ band centers or shapes at all longitudes. Similar to Ariel, ground-based and spacecraft spectra demonstrate that the 2.2-$\micron$ band is prevalent in the Pluto system, occurring on Pluto and its moons Charon, Nix, and Hydra \citep[e.g.,][]{brown2000evidence, cook2007near,grundy2016surface, cook2018composition, cruikshank2019recent,dalle2019detection}. The spectral signature of the 2.2-$\micron$ band on Pluto and its moons displays variations in its shape and central wavelength position, hinting at the presence of both NH$_3$-hydrates and NH$_4$-rich salts \citep[e.g.,][]{cook2018composition,protopapa2020charon}. Analogous to the Pluto system, perhaps the variable spectral signature of the 2.2-$\micron$ band detected on Ariel results from multiple contributing NH$_3$- and NH$_4$-bearing constituents, with different abundances at different locations across its surface. 

Although other constituents like CH$_4$ ice \citep[e.g.,][]{gerakines2005strengths}, complex organics \citep[e.g.,][]{cruikshank1991solid}, and hydrated silicates \citep[e.g.,][]{clark1990high} have prominent absorption features spanning the wavelength region of the 2.2-$\micron$ and 2.24-$\micron$ bands, these species are less likely to be the dominant contributors to these two bands. It is unlikely that CH$_4$ ice is currently present on Ariel's surface because it is highly volatile and should sublimate rapidly at Ariel's peak surface temperature ($\sim$80 - 90 K). Complex organics and hydrated silicates are much less volatile and could be present in the spectrally red material observed on the Uranian moons \citep{cartwright2018red}. This red material likely originated as dust particles liberated from the surfaces of Uranus' retrograde irregular satellites, which migrated inward due to Poynting-Robertson drag and mantled the leading hemispheres of the classical moons, in particular the outer moons Titania and Oberon \citep[e.g.,][]{tamayo2013chaotic}. However, the 2.2-$\micron$ and 2.24-$\micron$ bands are stronger on Ariel than on Titania and Oberon, and red material is relatively scarce on Ariel compared to the outer moons \citep{cartwright2018red}. Furthermore, the 2.2-$\micron$ and 2.24-$\micron$ bands are present over a wide range of longitudes on Ariel, spanning its leading and trailing hemispheres (Figure 1), unlike the distribution of red material, which is concentrated on the leading hemispheres of the Uranian moons. Therefore, NH$_3$-bearing constituents sourced from Ariel's interior are better candidate species to explain the presence of the 2.2-$\micron$ and 2.24-$\micron$ bands compared to complex organics or hydrated silicates delivered by dust impacts.  

Bombardment by charged particles trapped in Uranus' magnetosphere should efficiently decompose NH$_3$-rich deposits exposed on Ariel's surface in only $\sim$10$^6$ years \citep{moore2007ammonia}. Albeit, accurate models of moon-magnetosphere interactions at Uranus are lacking, and it is possible that charged particle weathering of NH$_3$-rich species is more efficient in some locations (i.e, on Ariel’s trailing hemisphere) compared to others (i.e., on Ariel’s leading hemisphere). Energetic protons, and ultraviolet (UV) solar photons,  are absorbed within the top 10 $\micron$ of H$_2$O ice-rich surfaces \citep[e.g.,][]{delitsky1998ice}. The surface of Ariel and the other Uranian moons could be mantled by a thin layer of small H$_2$O ice grains ($\sim$10 $\micron$ thick), with other constituents like CO$_2$ ice retained beneath this topmost layer \citep[e.g.,][]{cartwright2020probing}. The average NIR photon penetration depth into Ariel's regolith at 2.2 $\micron$ and 2.24 $\micron$ is $\sim$1.2 mm and $\sim$1.6 mm, respectively \citep[e.g.,][]{cartwright2020probing}. Therefore, NH$_3$-bearing species retained beneath the top $\sim$10 $\micron$ of Ariel's surface might not interact with UV photons or energetic protons. Energetic electrons ($\sim$1 MeV), however, can penetrate up to cm-scale depths into icy satellite regoliths \citep[e.g.,][]{nordheim2017near} and should readily interact with NH$_3$-bearing species and other constituents retained at depth. Furthermore, the flux of $\sim$1 MeV electrons is relatively high at the Uranian moons, comparable to the Jovian system \citep[e.g.,][]{mauk2010electron}. Nevertheless, an overlying veneer of H$_2$O ice might reduce the destruction rate of NH$_3$-rich deposits, and thereby increase their retention timescales on Ariel. 

Many irradiation-fragmented NH$_3$ molecules will either recombine back into NH$_3$ \citep[e.g.,][]{cruikshank2019recent} or form NH$_4$$^+$ ions \citep[e.g.,][]{moore2007ammonia}. These NH$_4$$^+$ ions could then interact with other constituents, including CO$_2$ ice, to form less volatile salts like (NH$_4$)$_2$CO$_3$, which would contribute to the 2.2-$\micron$ band.  Charged particle bombardment of NH$_3$, H$_2$O ice, and CO$_2$ ice could spur production of N-rich complex organics as well \citep[e.g.,][]{allamandola1988photochemical}, which would also contribute to the 2.2-$\micron$ band. Therefore, NH$_3$-bearing deposits retained at depth, and/or converted into other N-bearing species, could persist for longer periods of time than the estimated destruction rate for NH$_3$ exposed on Ariel's surface. 

Alternatively, NH$_3$-rich geologic landforms could be younger than the regional-scale crater density age estimates ($\sim$1 - 2 Ga), and thus, NH$_3$ might have been exposed more recently. A similar scenario exists on Pluto's moon Charon where young and fresh craters with bright ejecta deposits, like Organa, Nasreddin, Skywalker, and Candide, exhibit notably higher concentrations of NH$_3$ compared to the surrounding terrain \citep[e.g.,][]{grundy2016surface,protopapa2020charon}. NH$_3$ diffuses fairly rapidly through H$_2$O ice (4.0 x 10$^{-10}$ cm$^2$ s$^{-1}$ at 140 K, \citealt{livingston2002general}), and geologic processes that increase near-surface porosity and fracturing could increase this diffusion rate further. Consequently, geologic features formed by cryovolcanism, tectonism, impact events, and mass wasting might represent ideal conduits for the diffusion of NH$_3$ through H$_2$O ice-rich regoliths, thereby increasing the timescales over which the spectral signature of NH$_3$-bearing species persist in these landforms. However, the low surface temperatures of Ariel and the other Uranian moons (30 - 90 K) could substantially reduce the diffusion rate of NH$_3$ through H$_2$O ice, complicating this scenario. To more thoroughly investigate the possible connection between NH$_3$ and geologic landforms on Ariel, updated estimates of the retention timescales for NH$_3$ and surface age estimates, using local scale crater densities, are needed. Nevertheless, the spectral evidence presented here is consistent with Ariel experiencing geologic activity in the recent past, including possible emplacement of NH$_3$-rich cryolavas sourced from its interior. These results support the interpretation that Ariel is a possible ocean world, which has, or had, a global or regional subsurface liquid H$_2$O layer that communicated with its surface  \citep{hendrix2019nasa}.

Similar to Ariel, the four other classical Uranian moons display geologic landforms that suggest surface-interior communication via tectonism and cryovolcanism, in particular Miranda \citep[e.g.,][]{smith1986voyager,schenk1991fluid,beddingfield2015fault,beddingfield2020hidden}. NIR reflectance spectra of these four moons also display 2.2-$\micron$ bands \citep{bauer2002near,cartwright2018red}. Therefore, the presence of NH$_3$-rich constituents might have contributed to geologic activity on the other classical Uranian moons as well. Future telescope observations, made with available and proposed facilities, are needed to further investigate the spectral signature and spatial distribution of the 2.2-$\micron$ band on the Uranian moons \citep{cartwright2019exploring}. Data collected by an orbiting spacecraft in the Uranian system would revolutionize our understanding of these moons, providing spatially resolved datasets that would allow for unprecedented investigation of whether they are, or were, ocean worlds \citep{cartwright2020the}.

\section{Acknowledgments} 
This project was funded by NASA ROSES Solar System Observations grant NNX17AG15G and Solar System Workings grant NHH18ZDA001N. Laboratory measurements of amorphous NH$_3$ were funded by Solar System Workings grant 80NSSC18K00007. The observations reported here were made from the summit of Maunakea, and we thank the people of Hawaii for the opportunity to observe from this special mountain.

\bibliography{references}{}
\bibliographystyle{aasjournal}



\section{Appendix}

\subsection{Methods: Hapke-Mie spectral modeling procedure}
The radiative transfer models utilized in this study were generated using a hybrid Hapke-Mie approach, which has been applied previously to IRTF/SpeX datasets of these moons (Cartwright et al, 2015, 2018, 2020). This hybrid approach calculates the single scattering albedo ($\bar{\omega}$$_0$) for each constituent using Mie theory \citep[e.g.,][]{bohren1983light}. These $\bar{\omega}$$_0$ values are then utilized by Hapke equations to model the scattering properties of the observed regolith (e.g., Hapke, 2012). Mie theory models scattering and absorption by spherical particles that are distributed at random distances from one another. Mie theory can be used to model particles of any grain size. Consequently, $\bar{\omega}$$_0$ values calculated using Mie theory represent a valuable technique for generating synthetic spectra because many scattering models, including `pure' Hapke models, cannot generate robust results when considering constituents with grain diameters comparable to, or smaller than, the wavelength of incident light \citep[e.g.,][]{emery2006thermal}. To remove low amplitude resonance artifacts that can occur in scattering models that utilize Mie theory, our modeling software uses a small range of grain diameters ($\sim$10$\%$ spread) that are subsequently averaged together into the specified grain size for each constituent.

\subsection{Methods: Band parameter measurements}
\begin{figure}[h]
	\includegraphics[scale=0.9]{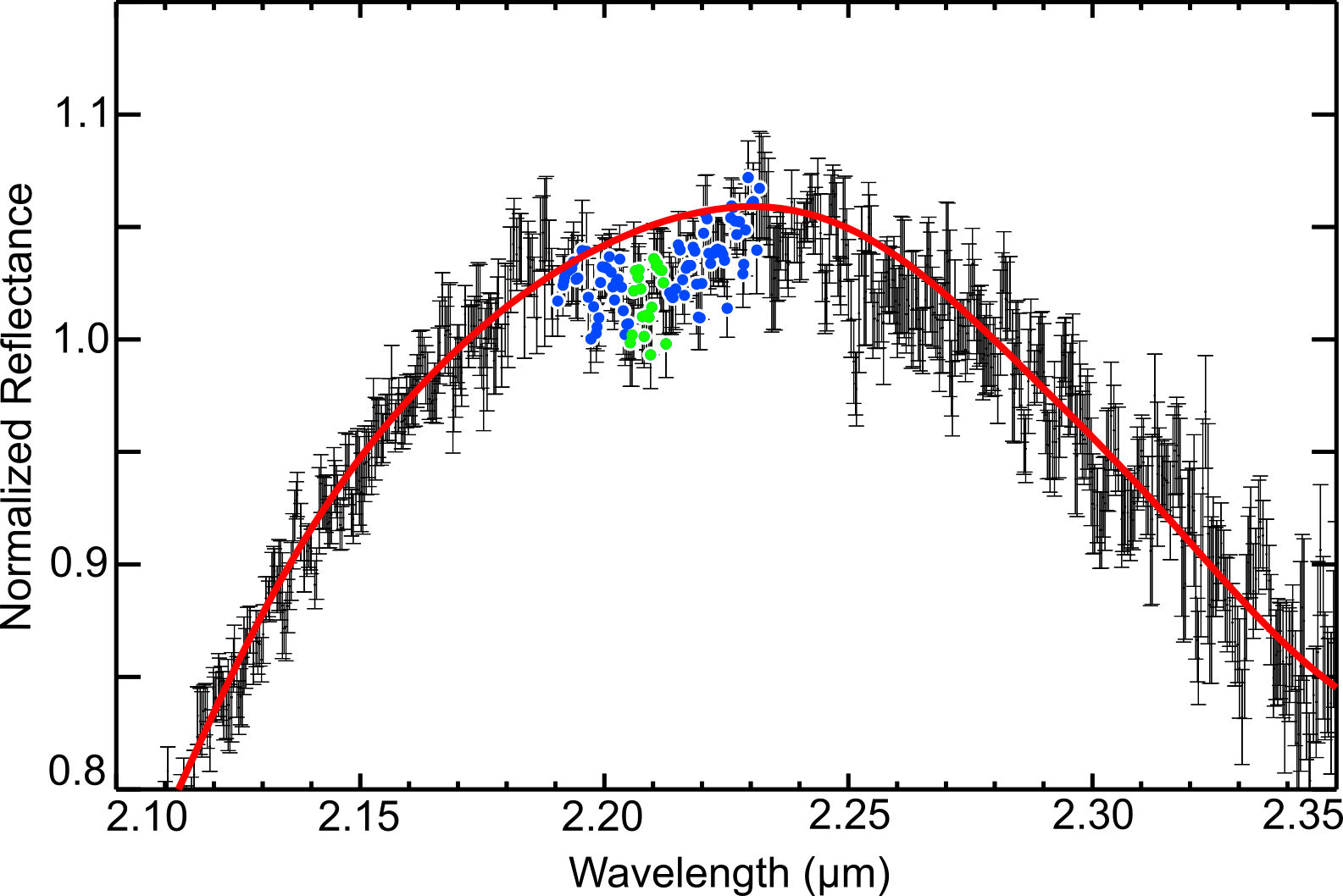}
	\caption{\textit{An example of our band area and depth measurement procedure, using the IRTF/SpeX spectrum of Ariel collected at a mid-observation longitude of 144.8$\degree$ (Spectrum 5 in Figures 1 and 3) (see Table 2 for 2.2-$\micron$ band measurement results). Data points used in the 2.2-$\micron$ band area measurement are highlighted (blue, spanning 2.190 to 2.232 $\micron$), as are the data points used to measure the band depth for this spectrum (green, centered at 2.209 $\micron$). The synthetic spectrum used to model the continuum (red line) is composed of: 86.3$\%$ 80 K crystalline H$_2$O ice (15 $\micron$ diameter grains), 0.39$\%$ 80 K crystalline H$_2$O ice (0.3 $\micron$ diameter grains), 3.8$\%$ amorphous carbon (9 $\micron$ diameter grains), and 9.51$\%$ CO$_2$ ice (5 $\micron$ diameter grains).}}\vspace{0.1 cm}
\end{figure} 

\clearpage

\subsection{Results: IRTF/SpeX spectra}

\begin{figure}[h!]
	\includegraphics[scale=0.99]{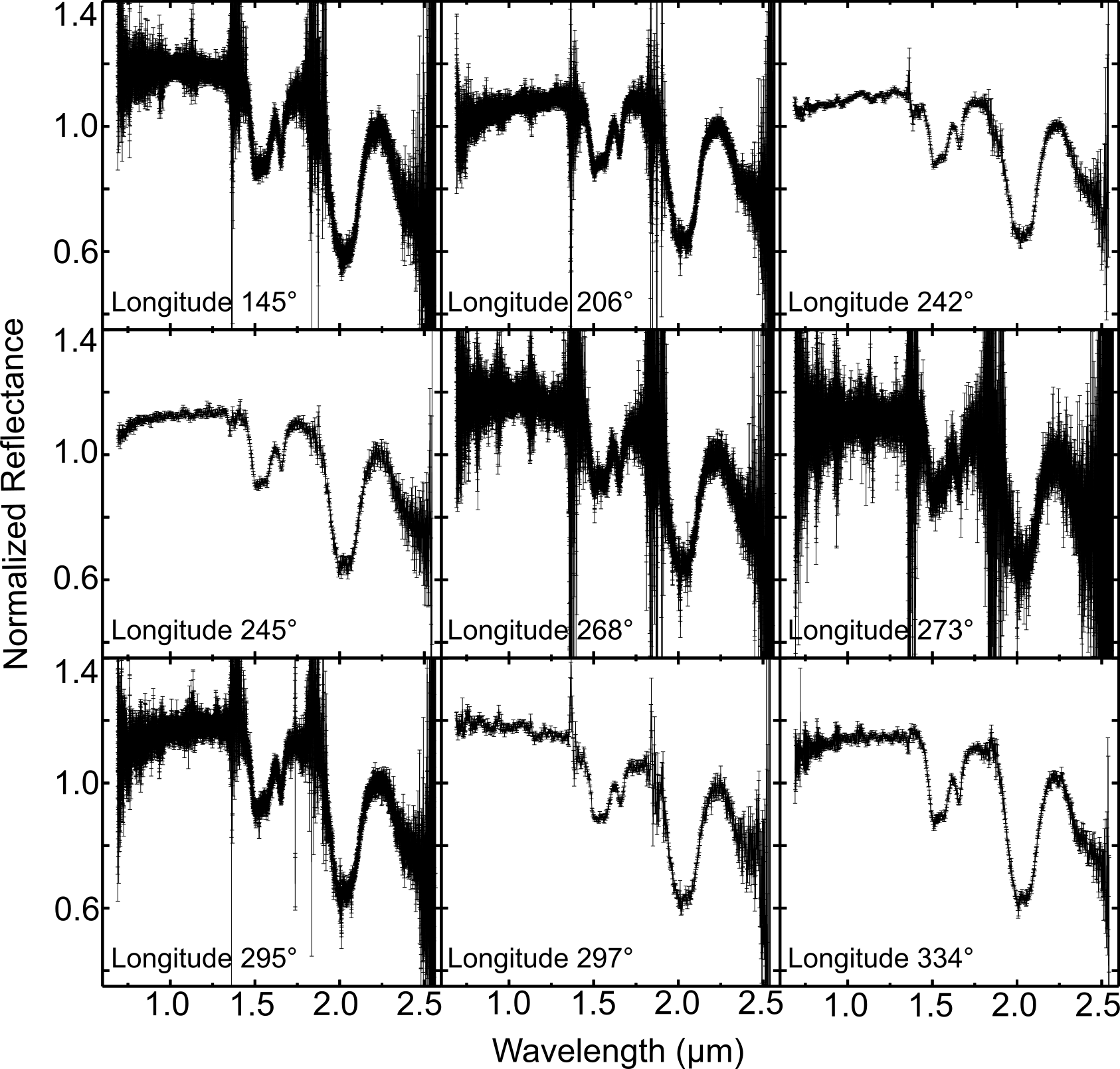}
	\caption{\textit{Nine new IRTF/SpeX spectra of Ariel and their 1$\sigma$ uncertainties, collected between 2017 and 2019. The mid-observation, sub-observer longitude for each spectrum is included in the bottom left-hand corner of each plot (see Table 1 for observation details). All spectra have been normalized to 1 between 2.24 to 2.25 $\micron$.}}\vspace{0.1 cm}
\end{figure} 

\subsection{Results: 2.24-$\micron$ band parameter measurements}

\begin{table}[hbt!]
	\caption {Measurements of the 2.24-$\micron$ band areas and depths.} 
	\hskip-2.0cm\begin{tabular}{llllllllllllll}
		\hline\hline
		\begin{tabular}[c]{@{}l@{}}\hspace{-1 cm}Sub-\\  \hspace{-1 cm}Observer \\ \hspace{-1 cm}Long. ($\degree$)\end{tabular} & \begin{tabular}[c]{@{}l@{}} \hspace{-1 cm}Sub-\\  \hspace{-1 cm}Observer \\  \hspace{-1 cm}Lat.  ($\degree$)\end{tabular}   & \begin{tabular}[c]{@{}l@{}} \hspace{-1 cm}Wavelength \\  \hspace{-1 cm}Range ($\micron$) \end{tabular} & \begin{tabular}[c]{@{}l@{}} \hspace{-1 cm}Band Area \\ \hspace{-1 cm}(10$^-$$^4$ $\micron$)   \end{tabular} & \begin{tabular}[c]{@{}l@{}} \hspace{-1 cm}Band Depth \\  \hspace{-1 cm}($\micron$)\end{tabular} & \begin{tabular}[c]{@{}l@{}} \hspace{-1 cm}$>$ 2$\sigma$ Band \\  \hspace{-1 cm}Area $\&$ Depth \\  \hspace{-1 cm}Measurement? \end{tabular} & \begin{tabular}[c]{@{}l@{}} \hspace{-1 cm}Band \\  \hspace{-1 cm}Center\\   \hspace{-1 cm}($\micron$) \end{tabular}\\
		\hline
		137.6 &  34.6 & 2.218 - 2.253 & 3.71 $\pm$ 1.22 & 0.023 $\pm$ 0.011 & Yes & 2.237 \\
		304.8 & -23.2 & 2.225 - 2.255 & 4.37 $\pm$ 1.85 & 0.033 $\pm$ 0.016 & Yes & 2.238 \\
		316.6 & -18.2 & 2.220 - 2.260 & 6.22 $\pm$ 1.55 & 0.037 $\pm$ 0.013 & Yes & 2.238  \\
		334.4 &  39.7 & 2.227 - 2.262 & 5.09 $\pm$ 0.86 & 0.028 $\pm$ 0.006 & Yes & 2.244  \\
		\hline
	\end{tabular}
\end{table}

\clearpage

\subsection{Results: Spatial distribution of the 2.2-$\micron$ band}
\begin{table}[h]
	\caption {\textit{F}-test analysis of the longitudinal distribution of the 2.2-$\micron$ band.} 
	\hskip-2.5cm\begin{tabular}{llllllllllllll}
		\hline\hline
		\begin{tabular}[c]{@{}l@{}}\hspace{-1 cm}Analyzed\\  \hspace{-1 cm}Measurement \end{tabular} & \begin{tabular}[c]{@{}l@{}} \hspace{-1 cm}One Tailed \\  \hspace{-1 cm}\textit{F}-test Ratio \end{tabular}  & \begin{tabular}[c]{@{}l@{}} \hspace{-1 cm}Sample \\  \hspace{-1 cm}Size (n) \end{tabular} &
		\begin{tabular}[c]{@{}l@{}} \hspace{-1 cm}Mean Model \\ \hspace{-1 cm}Degree of \\ \hspace{-1 cm}Freedom (n-1) \end{tabular} & \begin{tabular}[c]{@{}l@{}} \hspace{-1 cm} Sinusoidal \\ \hspace{-1 cm}Model Degree of \\ \hspace{-1 cm}Freedom (n-3) \end{tabular} & \begin{tabular}[c]{@{}l@{}} \hspace{-1 cm}Probability (\textit{p}) \\  \end{tabular} & \begin{tabular}[c]{@{}l@{}} \hspace{-1 cm}Reject Null  \\  \hspace{-1 cm}Hypothesis? \end{tabular}\\
		\hline
		2.2-$\micron$ Band Areas & 1.659 & 32 & 31 & 29 & 0.087 & No  \\
		2.2-$\micron$ Band Depths & 1.603 & 32 & 31 & 29 & 0.102 & No \\
		\hline
	\end{tabular}
\end{table}

\end{document}